\documentclass[10pt,aps,pra,twocolumn,showpacs]{revtex4-1}
\usepackage{graphicx}
\usepackage{bm}
\usepackage{amssymb,amsfonts,amsmath}
\usepackage{color}

\begin{document}

\newcommand{\etal}{\textsl{et al.}}
\newcommand{\GSA}{{ground state atom}}
\renewcommand{\vec}[1]{\bm{#1}}
\newcommand{\rr}{\vec{r}}
\newcommand{\RR}{\vec{R}}
\newcommand{\abs}[1]{\left|#1\right|}
\newcommand{\me}{m_\mathrm{e}}
\newcommand{\mRb}{m_\mathrm{Rb}}
\newcommand{\ee}{\mathrm{e}}
\newcommand{\dd}{\mathrm{d}}
\newcommand{\ii}{\mathrm{i}}
\renewcommand{\vec}[1]{\bm{#1}}
\newcommand{\vRb}{\vec{v}_\mathrm{Rb}}
\newcommand{\ve}{\vec{v}_\mathrm{e}}
\newcommand{\AZ}[1]{``#1''}
\newcommand{\psiryd}{\psi_\mathrm{Ry}}
\newcommand{\psipw}{\psi_\mathrm{pw}}
\newcommand{\order}[1]{\mathcal{O}\left(#1\right)}
\newcommand{\momentum}[1]{\vec{p}^{(#1)}}
\newcommand{\vbar}{\bar{v}}
\hyphenation{Ryd-berg}
\newcommand{\TR}[1]{\textcolor{blue}{#1}}

\title{Quantum-classical model for the formation of Rydberg molecules}
\author{Andrej Junginger, J\"org Main,  and G\"unter Wunner}
\affiliation{Institut f\"ur Theoretische Physik 1, Universit\"at
  Stuttgart, 70550 Stuttgart, Germany}
\date{\today}

\begin{abstract}
A fascinating aspect of Rydberg atoms is their ability to form huge
but very weakly bound molecules with a ground state atom, only held
together by a scattering process between the latter and the Rydberg
electron. Beyond the usual way of creating such molecules by laser
excitation from two \GSA s with a distance of less than the Rydberg
radius, we demonstrate that Rydberg molecules can also be formed by
capturing a \GSA\ which is initially located outside the range of the
Rydberg atom when it comes in contact with it. To demonstrate this
effect, we investigate the scattering process between the Rydberg
electron and the \GSA\ within a quantum-classical framework. In this
picture capturing results from a dissipative finite-mass correction
term in the classical equations of motion. We show that and under
which conditions the capturing takes place.
\end{abstract}

%
\maketitle
%
\section{Introduction}
Highly excited Rydberg atoms have long-since been in the focus of
numerous theoretical and experimental investigations
\cite{Bendkowsky2009, Bendkowsky2010, Greene2000, Saffman2010,
  Younge2009, Ates2007, Wuester2009, Butscher2011}, which exploit,
e.g., their large spatial extensions, which give rise to huge
polarizabilities and strong interactions, or their long lifetimes. In
2000, Greene \etal\ \cite{Greene2000} predicted that Rydberg atoms can
form very weakly bound molecules with \GSA s. This was experimentally
confirmed in 2009 by Bendkowsky \etal\ \cite{Bendkowsky2009} in a cold
dilute gas of rubidium atoms,  and  the results well agree with a
simple quantum mechanical model developed by Greene \etal\
\cite{Greene2000}.

Rydberg molecules are created by laser excitation of one \GSA\ in a
two-photon process using detuned lasers. The resulting Rydberg atom
can then form a molecule with another ground state atom, which, at the
time of laser excitation, is within a distance from the nucleus of the
Rydberg atom smaller than the extension of the wave function of the
Rydberg electron. In this way, also the creation of excited dimers and
Rydberg trimers is possible \cite{Bendkowsky2010, Butscher2011}.
The experiments reported in \cite{Butscher2011} also show the yet
unexplained formation of Rb$_2^+$ in the photoassociation spectra
without any detuning of the lasers, i.e., at resonance of the
Rydberg atom.

In this paper, we pursue a quantum-classical approach to understand
the mechanism by which Rydberg molecules are formed. In such a picture
we will demonstrate how ground state atoms which are initially located
at distances from the Rydberg atom larger than its extension can be
captured by the latter \emph{after} its excitation. This will lead to
the interpretation that a Rydberg molecule is formed in the sense of a
chemical reaction in which two reactants, the Rydberg and the \GSA ,
interact to form the reaction product, i.e.\ the Rydberg molecule. As
is well known chemical reactions cannot take place between two single
particles because energy and momentum cannot be conserved
simultaneously.  A reaction, however, can occur in the formation of a
Rydberg molecule because the Rydberg electron plays the role of a
third particle which can, within the natural line width of the Rydberg
state, absorb kinetic energy from the ground state atom.

Our paper is organized as follows: First, we give a brief review and
discussion of Greene's quantum mechanical model before we introduce
our quantum-classical treatment. The latter is based on classically
describing the motion of the Rydberg electron in the $1/r$-potential,
locally approximating it as a superposition of plane waves at the
position of the \GSA , and quantum mechanically describing the
scattering process. As a result, we can derive an expression for the
force acting on the \GSA\ which includes a dissipative finite-mass
correction term in the classical equations of motion. Finally, we
compare the results obtained for  the models with and without the
correction term, and determine under which conditions the capturing of
a \GSA, i.e., the formation of a molecule, can occur.

\section{Theory}
As a framework for the theoretical description of the dynamics one could
use Quantum Molecular Dynamics \cite{Bornemann1996} classically
describing the motion of the nuclei of the molecule under the
influence of the electron. However, in the case of Rydberg molecules,
the dynamics of the Rydberg electron is predominantly determined by
the nucleus of the Rydberg atom since the latter is positively
charged, and the \GSA\ is a neutral particle which only acts as a
small perturber. Thus, the wave function of the Rydberg electron is
known for given quantum numbers and an appropriate description of the
interaction between the Rydberg electron and the \GSA\ in a
quantum-classical picture is therefore possible by \emph{locally}
approximating the known Rydberg wave function by a superposition of
plane waves, each representing one corresponding classical Kepler
ellipse of the Rydberg electron.

The following two sections review the quantum mechanical description
of Rydberg atoms and introduce the quantum-classical treatment of the
scattering process.

\subsection{Molecular potential of Rydberg molecules}
\begin{figure}[t]
\includegraphics[width=\columnwidth]{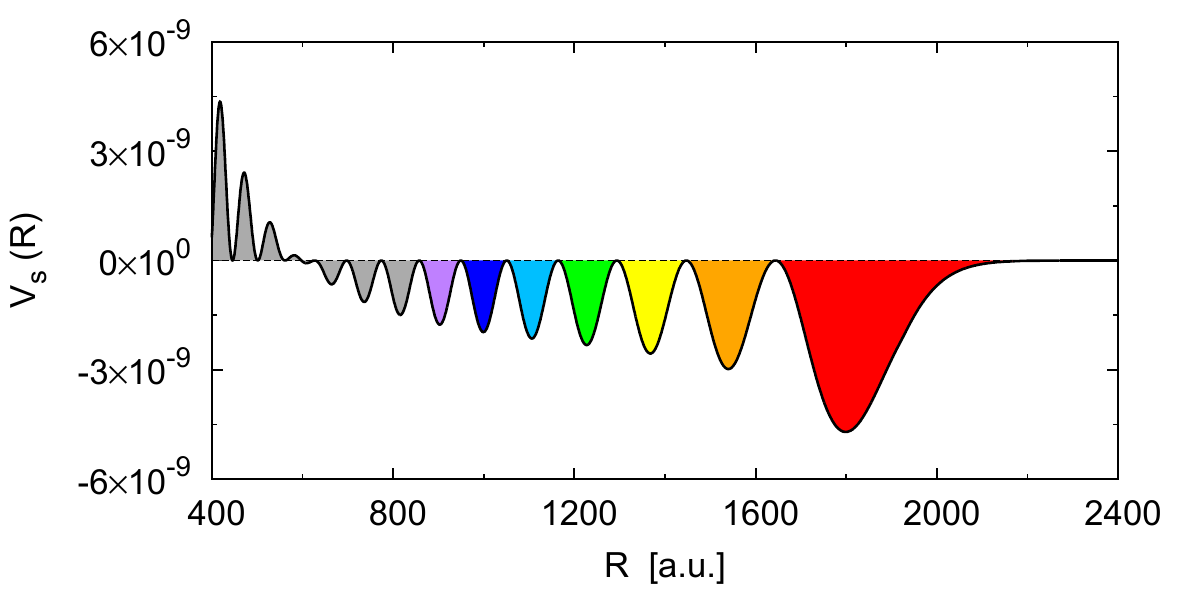}
\caption{\label{fig-potential}
          Molecular potential of a rubidium Rydberg molecule as a
          function of the internuclear distance for a negative
          scattering length $a_0=-18.5$ and a polarizability
          $\alpha=319.0$ in Eq.\ (\ref{eq-scattering-length}),
          calculated from Eq.\ (\ref{eq-V-s}) for quantum
          numbers $n=31$, $l=0$, $m=0$. The assignment of colors to
          the different potential wells will be the same in  Fig.\
          \ref{fig-capturing}.}
\end{figure}

In the quantum mechanical model of Greene \etal\ \cite{Greene2000}
the interaction between the Rydberg electron and the \GSA\ is
described using a Fermi type pseudopotential \cite{Fermi1934} 
\begin{equation}
	V(\rr,\RR) = 2 \pi a_\text{s}(k) \delta(\rr-\RR),
	\label{eq-Fermi-pseudopotential}
\end{equation} where $\rr$ and $\RR$ denote the positions of the
Rydberg electron and the \GSA, respectively, and the function $a_\text{s}(k)$
is the s-wave scattering length, which depends on the wave vector
$k$ of the Rydberg electron. In a first-order approximation, it can be
expressed in the form \cite{Omont1977}
\begin{equation}
	a_\text{s}(k) = a_\text{s,0} + \frac{\pi}{3} \alpha k + \order{k^2}.
	\label{eq-scattering-length}
\end{equation} 
Here $a_\text{s,0}=-18.5\,$a.u.\ is the zero-energy scattering length and
$\alpha=319$ the polarizability of the target \cite{Bendkowsky2009, Molof1974}. Within a mean-field
approximation this contact interaction leads to the molecular
potential
\begin{equation}
	V_\text{s}(\RR) = 2 \pi a_\text{s}(k) \abs{\psiryd(\RR)}^2,
	\label{eq-V-s}
\end{equation} where $\psiryd (\RR)$ is the value of the wave function
of the Rydberg electron at the position $\RR$ of the \GSA. If $a_\text{s}(k) <
0$ the interaction is attractive. Fig.\ \ref{fig-potential} shows the
molecular potential \eqref{eq-V-s} for a particular set of quantum numbers.

In the potential shown in Fig.\ \ref{fig-potential} and used in
Refs.\ \cite{Bendkowsky2009,Greene2000} only s-wave scattering is
taken into account. A more realistic potential is obtained when also
p-wave scattering is taken into account by the additional term
\begin{equation}
 V_\text{p}(\RR) = 6 \pi a_\text{p}^3 \abs{\nabla \psi_{\rm Ry}(\RR)}^2
\label{eq-V-p}
\end{equation}
with $a_\text{p}=-21.15\,$a.u.\ \cite{Bendkowsky2010} so that the molecular 
potential is given by
\begin{equation}
V(\RR) = 
\begin{cases}
  V_\text{s}(\RR), & \text{pure s-wave scat.} \\
  V_\text{s}(\RR) + V_\text{p}(\RR), & \text{with p-wave scat.}
\end{cases}%
\label{eq-potential}%
\end{equation}%
As will be shown in
Sec.\ \ref{sec:results} the occurrence of the quasi-classical formation of Rydberg
molecules does not depend on whether or not p-wave scattering is
considered.

By construction, Eq.\ (\ref{eq-potential}) associates a \emph{fixed}
position $\RR$ of the \GSA\ with the potential energy $V(\RR)$.  If
one considers a single scattering event of the Rydberg electron with
the \GSA\ this is physically equivalent to the assumption that the
center of mass of the two scattering partners lies exactly at the
center of the \GSA, i.e.\ $\me/m_\mathrm{gs} = 0$. Moreover, Eq.\
(\ref{eq-potential}) only takes into account the mean density
distribution of the Rydberg electron and thus neglects dynamical
effects of the scattering process.

\subsection{Scattering process within the quantum-classical framework}
The assumption $\me/m_\mathrm{gs} = 0$ is, of course, never strictly
fulfilled, and, e.g., for rubidium atoms, which we consider
throughout this paper, we have $\me/\mRb \approx 6\times
10^{-6}$. Thus the description of the electron-Rb-scattering in the
framework of Eq.\ (\ref{eq-potential}) is not complete. To take into
account dynamical effects of single scattering events, we will
describe these in a quantum-classical way. In the classical equations
of motion, by which we describe the dynamics of the heavy \GSA , this
treatment will lead to a small but nonvanishing dissipative correction
term in the order of the ratio of the masses of the two scattering
partners, $\order{\me/\mRb}$. In spite of its smallness it can have
drastic effects on the dynamics of the rubidium atoms, as we will
demonstrate in Sec.\ \ref{sec:results}.

Our quantum-classical model is based on the following assumptions: We
describe the Rydberg atom as hydrogen-like with one Rydberg electron
and a core with charge $+e$. Because of the high excitation of the
Rydberg atom ($n\gg 1$), correspondence principle allows us to treat
the motion of the Rydberg electron in terms of the classical
trajectories, namely Kepler ellipses, and we quantize the latters'
angular momentum $L$ and their energy $E$ semiclassically according to
\begin{equation}
	L = l+\frac{1}{2},\qquad E = \frac{\vec{p}^2}{2} - \frac{1}{r}
        = -\frac{1}{2n^2},
\end{equation}
where $n=1,2,3,\ldots$ and $l=0,1,2,\ldots$ are the principal and
angular momentum quantum numbers, respectively, $\vec{p}$ is the momentum of
the Rydberg electron and $r=\abs{\rr}$ is its distance from the
core. Note that the continuous set of Kepler ellipses fulfilling these
conditions only differ from each other by a rotation of the ellipses
around the azimuthal quantization axis.

\begin{figure}[t]
	\includegraphics[width=\columnwidth]{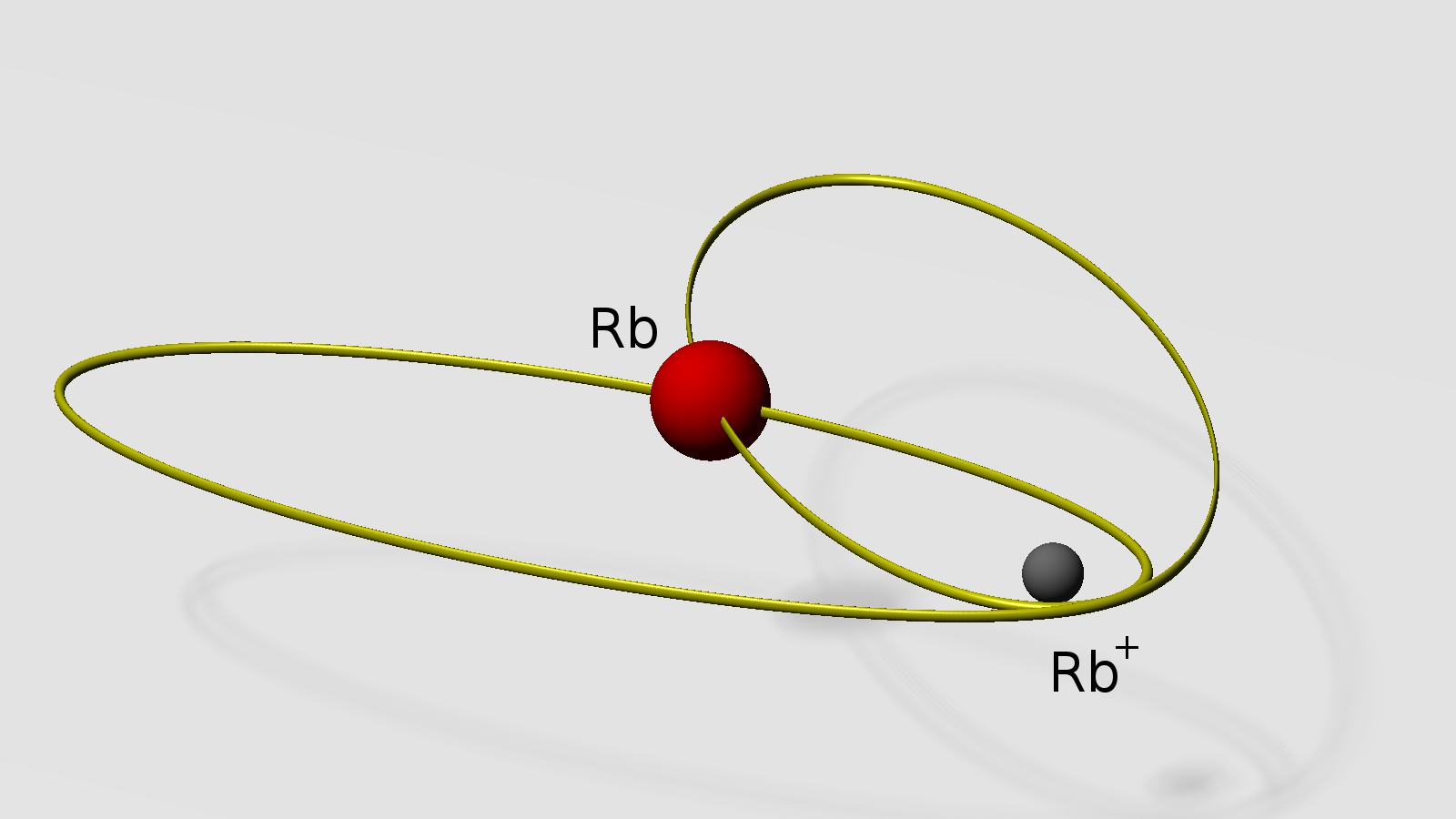}
	\caption{\label{fig-scheme}
Schematic drawing of the quantum-classical model for
          the interaction between the Rydberg electron, orbiting the
          Rb$^+$ core (small gray sphere), and the Rb \GSA\ (big red
          sphere). For  the magnetic quantum number $m=0$, considered
          here, there are only two Kepler orbits on which the Rydberg
          electron can hit the \GSA\, and each is traversed in
          clockwise and counterclockwise direction.}
\end{figure}

Since the interaction between the Rydberg electron and the \GSA\ is of
contact-like type, within this quantum-classical framework these two
will only interact if the Kepler ellipses hit the latter (see Fig.\
\ref{fig-scheme}), i.e.\ if the orbit includes the position $\RR$ of
the \GSA.
It can be easily shown that out of the infinite set of
equivalent Kepler ellipses with given quantum numbers $n, ~l, ~m$ only
\emph{four} ellipses fulfill this additional condition (see also Ref.\
\cite{Granger2001}). In experiments only molecules with the Rydberg
atom in an s-state have been formed so far, therefore we will also
restrict ourselves to the angular momentum quantum numbers $l = m =
0$. In this case always two of the four possible ellipses coincide,
and we are left with only two Kepler orbits that can intersect with
the \GSA\, each of which is traversed in clockwise and
counterclockwise direction. Thus the momenta $\momentum{i}$ of the
Rydberg electron on the $i$-th Kepler ellipse at the point of
intersection are opposite to each other,
\begin{equation}
 \momentum{1} = -\momentum{2}, \qquad \momentum{3} = -\momentum{4}
\label{eq-momenta-relation}
\end{equation}
for $m=0$. Their values can be easily calculated for
each point on the orbit and, in particular, at the point of
intersection, i.e., the position of the \GSA .

To describe the process of the Rydberg electron orbiting on a Kepler
ellipse being s-wave scattered at the \GSA\ we make use of the fact
that the extension of the highly excited Rydberg atom is, by far,
larger than that of the \GSA. Thus, we can assume the Coulomb
potential $V_\text{C}(\rr)$ of the Rydberg atom's nucleus to be constant in a
small vicinity of the \GSA , i.e.\ $V_\text{C}(\rr)\approx \mathrm{const.}$
for $\rr\approx\RR$, further allowing us to approximate the Rydberg
electron on the $i$-th Kepler ellipse by a plane wave $\psipw^{(i)}
(\rr) = A^{(i)} \exp\left( \ii \vec{p}^{(i)} \vec{r} \right)$. The
\emph{local} approximation of the total wave function is consequently
a superposition of four plane waves, corresponding to the four
ellipses,
\begin{equation}
	\psiryd (\rr) \approx \sum_{i=1}^{4}	\psipw^{(i)} (\rr) =
        \sum_{i=1}^{4} A^{(i)} \exp\left( \ii \vec{p}^{(i)} \vec{r}
        \right),
\end{equation} which establishes the key link between the classical
picture of the interaction and the quantum mechanical description of
the scattering process: The momenta are identical to those of the
electron on the Kepler orbit at the point of collision, and the
amplitudes $A^{(i)}$ are determined by fitting the wave function to
the exact quantum mechanical Rydberg wave function. For $m=0$ this
wave function is real-valued and cylindrically symmetric, $\psiryd
=\psiryd(\rho, z)$, which implies that the complex amplitudes
$A^{(i)}$ come in complex conjugate pairs, $A^{(1)}=A^{*(2)}$ and
$A^{(3)}=A^{*(4)}$. We are therefore left with four unknowns, the real
and imaginary parts of $A^{(1)}$ and $A^{(3)}$. To determine these we
require that at the intersection point the values of the wave
functions and their first derivatives coincide:
\begin{subequations}
\begin{align}
  \psipw |_{\rr=\RR} &= \psiryd |_{\rr=\RR}\;,
\label{eq-system-of-equations-a}\\
  \partial_\rho \psipw |_{\rr=\RR} &= \partial_\rho \psiryd |_{\rr=\RR}\;,
\label{eq-system-of-equations-b}\\
  \partial_z \psipw |_{\rr=\RR} &= \partial_z \psiryd |_{\rr=\RR}\;.
\label{eq-system-of-equations-c}	
\end{align}
\label{eq-system-of-equations}
\end{subequations}
Requiring also the identity of the second derivatives would provide
three more equations but render the total set of equations
over-determined. To obtain a fourth equation we therefore only require
that the sum of the moduli squared of the deviations of the second
derivatives of the quantum-classical and the Rydberg wave function be
a minimum
\begin{multline}
 \left(\partial_\rho^2(\psipw - \psiryd)|_{\rr=\RR}\right)^2 
 + \left(\partial_z^2(\psipw - \psiryd)|_{\rr=\RR}\right)^2 \\
 + \left(\partial_{\rho,z}^2(\psipw - \psiryd)|_{\rr=\RR}\right)^2 = \min.
\label{sec:derivs} 
\end{multline}
This leads to the best possible approximation of the
Rydberg wave function by the four plane waves $\psipw^{(i)}$.

To describe the scattering process in the quantum-classical
picture, we first consider scattering of a \emph{single} electron
with a \GSA\ in the latter's rest frame. The incoming 
Rydberg electron on the $i$-th Kepler ellipse, described
by the plane wave $\psipw^{(i)}$ with momentum
$\vec{p}_\mathrm{in}^{(i)} = \me \ve^{(i)} = \hbar \vec{k}^{(i)}$, is scattered to an
outgoing wave
\begin{equation}
  \psi_\mathrm{out}^{(i)} \sim \frac{\exp\bigl( \ii\,
    {p}_\mathrm{out}^{(i)} \abs{\rr-\RR} \bigr)}{\abs{\rr-\RR} } \,
  \left(\,f_\text{s} + f_\text{p} \cos \theta  + \ldots \, \right)
  \label{eq-psi-out}
\end{equation}
where $f_l$ are the scattering amplitudes that are related to the scattering phase shifts $\delta_l$ by $f_l = k^{-1}(2l+1) \, \ee^{\ii \delta_l} \sin \delta_l$.

The wave number of the Rydberg electron is given by $k = \sqrt{2/R - 1/n^2}$ so that we obtain $k\lesssim 0.025$ for typical values $n=31$ and $R\gtrsim 1200$ (see below). For rubidium, scattering in this region is therefore dominated by the s-wave so that $\delta_\text{p}$ and contributions from higher partial waves can be neglected in the scattering process.

Because of the spherically symmetric angle distribution the total momentum of the outgoing s-wave is $\vec{p}_\mathrm{out}^{(i)}=0$ so that the Rydberg electron transfers a momentum of $\Delta
\vec{P}_\mathrm{Rb}^{(i)} = \me \ve^{(i)}$ to the target, i.e.\ the
rubidium \GSA. A number of $N^{(i)}$ colliding electrons consequently leads to a
momentum transfer of
\begin{equation}
  \Delta \vec{P}_\mathrm{Rb}^{(i)} = N^{(i)} \me \ve^{(i)} ,
\label{eq-momentum-transfer}
\end{equation} and if the scattering events occur in a time $\Delta t$
this corresponds to a classical force
\begin{equation}
 \vec{F}_\mathrm{Rb}^{(i)} = \Delta \vec{P}_\mathrm{Rb}^{(i)}/{\Delta t}
\end{equation}
acting on the \GSA . We now proceed from single but
continuous scattering processes to a current density
\begin{equation}
 \vec{j}_\text{e}^{(i)} = n_\mathrm{e}^{(i)} \vec{v}_\text{e}^{(i)} =
 \frac{N^{(i)}}{\sigma \Delta t} \, \hat{\vec{e}}_{\ve^{(i)}} 
\label{eq-current density}
\end{equation}
where $n_\mathrm{e}^{(i)}=\abs{A^{(i)}}^2$ is the
electron density on the $i$-th Kepler ellipse, $\sigma=4\pi a_\text{s}^2(k)$ is
the scattering cross-section, and $\hat{\vec{e}}_{\ve^{(i)}}$
is the unit vector in the direction of $\ve^{(i)}$. Similar to the discussion above, including p-wave scattering does not change the total scattering cross section $\sigma$ significantly in the important region $R \gtrsim 1200\,$a.u. Combining Eqs.\
(\ref{eq-momentum-transfer})--(\ref{eq-current density}) we end up
with
\begin{equation}
  \vec{F}_\mathrm{Rb}^{(i)} = n_\mathrm{e}^{(i)} \me \sigma
  \abs{\ve^{(i)}}^2  \hat{\vec{e}}_{\ve^{(i)}}.
\label{eq-force-restframe}
\end{equation} 

We now switch to the laboratory frame, where the \GSA , in general,
moves with a velocity $\vRb \neq 0$ relative to the ionic core of the
Rydberg atom and, without loss of generality, assume the latter to be
at rest. The transformation to the laboratory frame then results in
the formal substitution $\ve \to \ve - \vRb$ in Eq.\
(\ref{eq-force-restframe}), which leads to
\begin{subequations}
\begin{align}
  \vec{F}_\mathrm{Rb} &= \sum_{i=1}^4 \vec{F}_\mathrm{Rb}^{(i)},  \\
  \vec{F}_\mathrm{Rb}^{(i)} &= n_\mathrm{e}^{(i)} \me \sigma
      \abs{\vec{v}_\text{e}^{(i)} - \vec{v}_\mathrm{Rb}}^2
      \hat{\vec{e}}_{\left(\vec{v}_\text{e}^{(i)} - \vec{v}_\mathrm{Rb}\right)}
\end{align}
\label{eq-force}
\end{subequations}
Note that by locally describing the four Kepler ellipses as
\emph{independent} plane waves we lose 
all interference terms and thus the nodal structure of the Rydberg
wave function. Therefore the Eqs.\ (\ref{eq-force}) cannot catch the
mean electron density distribution. Moreover, since
$\vec{F}_\mathrm{Rb}^{(i)} \sim n_\mathrm{e}^{(i)}$ with the electron
density $n_\mathrm{e}^{(i)} = |\psi_\mathrm{pw}^{(i)}(\rr)|^2$, any
phase-factor corrections which may occur when transforming from the
\GSA 's rest frame to the laboratory frame will cancel out.

For a simple discussion of the effect of the forces acting on the
\GSA\ described by the Eqs.\ (\ref{eq-force})  imagine the interaction
with only the two coinciding Kepler ellipses $i=1,2$ (the same holds,
of course, for $i=3,4$) and the case where at the collision the \GSA\
and the Rydberg electron fly in the same direction,
$\hat{\vec{e}}_{\vec{v}_\mathrm{Rb}} =
\hat{\vec{e}}_{\vec{v}_\text{e}^{(1)}}$. Since Eq.\
(\ref{eq-momenta-relation}) then implies
$\hat{\vec{e}}_{\vec{v}_\text{e}^{(2)}}= -\hat{\vec{e}}_{\vec{v}_\text{e}^{(1)}}$,
we obtain a net force on the \GSA
\begin{equation}
  \vec{F}_\mathrm{Rb} = \vec{F}_\mathrm{Rb}^{(1)} +
  \vec{F}_\mathrm{Rb}^{(2)} = - 4 \me \sigma
  n_\mathrm{e}^{(1,2)} v_\mathrm{e} ^{(1,2)} v_\mathrm{Rb}
  \hat{\vec{e}}_{\vec{v}_\mathrm{Rb}}
\label{eq-force-example}
\end{equation} 
which for any value of the modulus of its velocity
$v_\mathrm{Rb}>0$ is directed opposite to its direction of flight,
i.e., the \GSA\ is decelerated. Note that in Eq.\
(\ref{eq-force-example}) always one contribution to the force is
accelerating (here the one with $\hat{\vec{e}}_{\vec{v}_\mathrm{Rb}} =
\hat{\vec{e}}_{\vec{v}_\text{e}^{(1)}}$) and the other one is decelerating
(here the one with $\hat{\vec{e}}_{\vec{v}_\mathrm{Rb}} =
-\hat{\vec{e}}_{\vec{v}_\text{e}^{(2)}}$), whereas the latter dominates since
$|\vec{v}_\text{e}^{(1)} - \vec{v}_\mathrm{Rb} | < |\vec{v}_\text{e}^{(2)} -
\vec{v}_\mathrm{Rb} |$. The discussion can be generalized to the case
of arbitrary flight directions $\hat{\vec{e}}_{\vec{v}_\mathrm{Rb}}$,
and there always occurs a deceleration of the \GSA\ as a net effect.

Taking into account both the potential Eq.\ (\ref{eq-potential})
resulting from the mean electron density distribution and the
dissipative correction terms due to the dynamical effects, Eqs.\
(\ref{eq-force}), one can write down the classical equations of motion
for the dynamics of the \GSA\ under the influence of the Rydberg atom:
\begin{multline}
	\frac{\dd^2 \RR}{\dd t^2} = -\frac{1}{\mRb} \vec{\nabla} V(\RR) \\ 
	+ \frac{\me}{\mRb} \sigma \sum_{i=1}^4 n_\mathrm{e}^{(i)}
        \abs{\ve^{(i)} - \vRb}^2 \hat{\vec{e}}_{\ve^{(i)} - \vRb} \;. 
\label{eq-DGL}
\end{multline}
The second term is on the order of $\me/\mRb \approx 6\times 10^{-6}$
and therefore small but, as will be shown below, can have drastic
effects on the motion of the \GSA . Note that in the limit $\me/\mRb
\to 0$ we recover the original model of  Greene \etal\ \cite{Greene2000}.

\section{Results and discussion}
\label{sec:results}
We obtain the following results solving the differential equation
(\ref{eq-DGL}) for initial values $\RR$ and $\vRb$. The physical
parameters are chosen in such a way that they cover the experiment in
Ref.\ \cite{Bendkowsky2009}, in which $^{87}$Rb atoms have been
excited to Rydberg s-states $n'\geq34$. Including quantum defect
corrections of $\delta=3$ for rubidium \cite{Li2003}, we therefore use
quantum numbers $n=n'-\delta=31$, $l=0$, and $m=0$.

Since we assume the Rydberg atom to be in the spherically symmetrical
s-state, there are only three quantities which determine the
dynamics of the \GSA, namely the absolute value of the initial
velocity of the \GSA, $v_\mathrm{Rb}$, the initial internuclear
distance $R_0$  and the angle $\phi$ between the direction of the
initial velocity and the line connecting the ground state atom and the
center of the Rydberg atom, $\cos \phi = \vRb \RR/(v_\mathrm{Rb}
R)$. Throughout this section we choose $R_0=2500\,$a.u.\ which is
outside the range of the Rydberg electron including the tail of the
Fermi pseudopotential.
In the following Sec.\ \ref{sec:swave} we first present the
results for the quasi-classical model using s-wave scattering. The
modifications obtained by including p-wave scattering are discussed
in Sec.\ \ref{sec:pwave}. An interpretation of the model beyond the
quasi-classical approach is outlined in Sec.\
\ref{sec:beyond-quantum-classical}.

\subsection{Model with s-wave scattering}
\label{sec:swave}
\begin{figure}[t]
\includegraphics[width=\columnwidth]{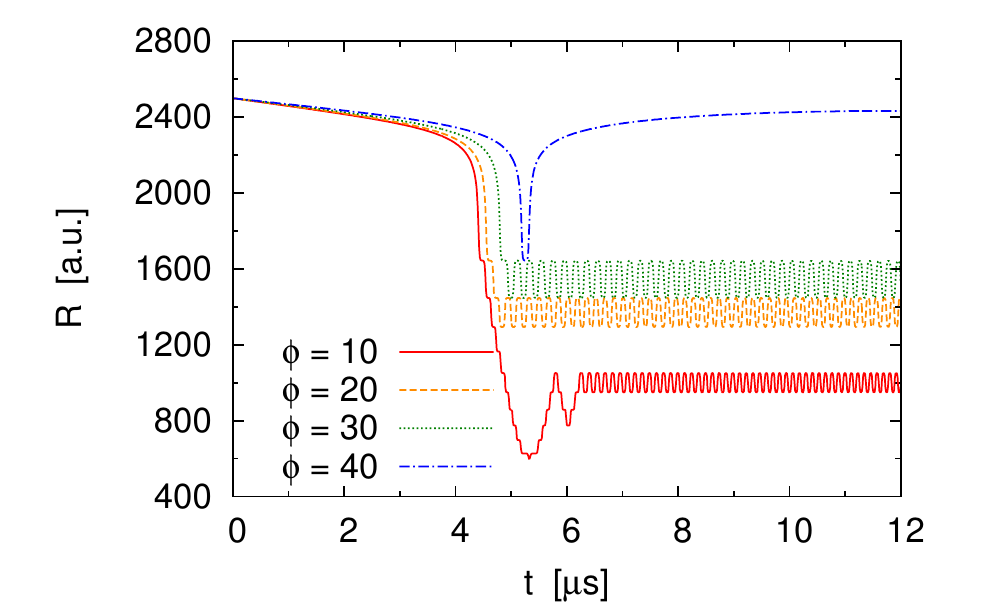}
\caption{\label{fig-angles}
 Internuclear distance between the Rydberg and the \GSA\ in dependence
 of time for different angles $\phi$ between the initial direction of
 the motion of the \GSA\ and the connecting line to the nucleus of the
 Rydberg atom. The initial velocity is set to $v_\mathrm{Rb}=
 10^{-9}\,$a.u. For all angles shown the \GSA\ will be captured but the
 selection of a specific local minimum strongly depends on the initial
 value of $\phi$.}
\end{figure}

To demonstrate the effect of the finite-mass correction term in Eq.\
(\ref{eq-DGL}) we discuss the trajectories of the \GSA\ obtained with
the correction. Fig.\ \ref{fig-angles} shows the internuclear distance
between the Rydberg and the \GSA\ for initial values
$v_\mathrm{Rb}= 10^{-9}\,$a.u.\ and various angles between
$\phi=10$ and $\phi=40$ degrees.

If only the conservative potential is taken into account, the \GSA\ approaches
the Rydberg atom, reaches some minimum distance and then leaves it
again. 
Taking into account the finite-mass correction term, the
\GSA\ initially shows a similar behavior, but on its way out from the
Rydberg atom, it begins to oscillate, e.g., around a distance of
$R\approx 1500\,$a.u.\ for the orbit launched with $\phi=30^\circ$ in
Fig.\ \ref{fig-angles}, and no longer leaves the Rydberg atom. The
physical meaning is that the \GSA\ has been captured in one of the
potential wells, i.e., the total energy of the \GSA\ has decreased
below zero and it cannot leave the Rydberg atom any more:
\begin{equation}
	E_\mathrm{g.a.} = \frac{\mRb}{2}\vRb^2 + V(\RR) < 0.
\label{eq-energy-below-zero}
\end{equation} 
We emphasize that the formation of the molecule occurs \emph{after}
the excitation of the Rydberg atom and not, as usual, by excitation
with a detuned laser.

In the example with $\phi=10$ degrees shown in Fig.\ \ref{fig-angles}
the \GSA\ enters the vicinity of the Rydberg atom at $t \approx
3\,\mu$s. The total energy $E_\mathrm{g.a.}$ of the ground state atom
becomes negative at $t \approx 5.5\,\mu$s, which means that the
Rydberg molecule is formed within about $2.5\,\mu$s. Note that this
time is significantly smaller than the lifetime of the Rydberg
molecule of about $15\,\mu$s \cite{Bendkowsky2009}.

Since the system of the ionic Rydberg core, the Rydberg electron, and
the \GSA\ is closed, an important issue of the process described here
is conservation of energy. In order to observe a deceleration of the
\GSA , its initial kinetic energy has to be transferred to its
scattering partner, i.e.\ the Rydberg electron. To satisfy Eq.\
(\ref{eq-energy-below-zero}) for the initial velocity
$v_\mathrm{Rb}= 10^{-9}\,$a.u., in total, a kinetic energy of
$E_\mathrm{kin}=525\,$Hz has to be absorbed. However, this amount of
energy is small compared to the level spacing between the quantum
state considered and a neighboring one, so that the excitation of a
higher quantum state would be highly off-resonant and, thus, quantum
mechanically forbidden. Nevertheless, the process remains allowed due
to the very exotic conditions in the ultra-cold Rydberg gas: The
quantum state $n$ of the Rydberg electron is not sharp, but has a
natural line width of $\Gamma/2\pi \approx 21\,$kHz resulting from the
$15\,\mu$s lifetime of the Rydberg molecule. The extremely small
amount of kinetic energy will, therefore, \emph{not} be deposited into
an excitation of a higher quantum state but is absorbed \emph{at
resonance} within the line width of the same state.

After the point of capturing, the classical computations show a
further decrease of the \GSA 's energy, which is, within the lifetime
of the Rydberg molecule, however, small compared to the depth of the
potential and thus the molecule will have dissociated long before
reaching binding energies of the quantized stationary vibrational
states of the molecule. In addition, the latter has to be excluded
because of physical reasons, since the required amount of energy
cannot be deposited in the excitation of the Rydberg electron within
the natural line width of the Rydberg state.

Moreover, we find a crucial dependence of the minimum of the potential
in which the \GSA\ is captured on the initial angle $\phi$ (see Fig.\
\ref{fig-angles} for some exemplary trajectories). In fact, for each
potential well there is a corresponding range of angles $\phi$ leading
to the formation of a molecule with an internuclear distance
associated to the particular well.

To generally determine for which initial conditions the \GSA\ will be
captured by the Rydberg atom, we calculate collisions for angles
$\phi$ between $0^\circ$ and $90^\circ$  and initial velocities
$v_\mathrm{Rb}$ from $10^{-9}\,$a.u.\ to 
$10^{-8}\,$a.u.\ corresponding to kinetic energies of about $525\,$Hz to
$52.5\,$kHz. As a reference for the velocity we take the mean velocity
$\vbar \approx 1.3\times 10^{-8}\,$a.u.\ of an ideal gas at the
temperature of $T=3.5\,\mu\mathrm{K}$ at which the experiment of
Bendkowsky \etal\ \cite{Bendkowsky2009} was performed, i.e., this
range of velocity corresponds to \AZ{slow} \GSA s. 

\begin{figure}[t]
\includegraphics[width=\columnwidth]{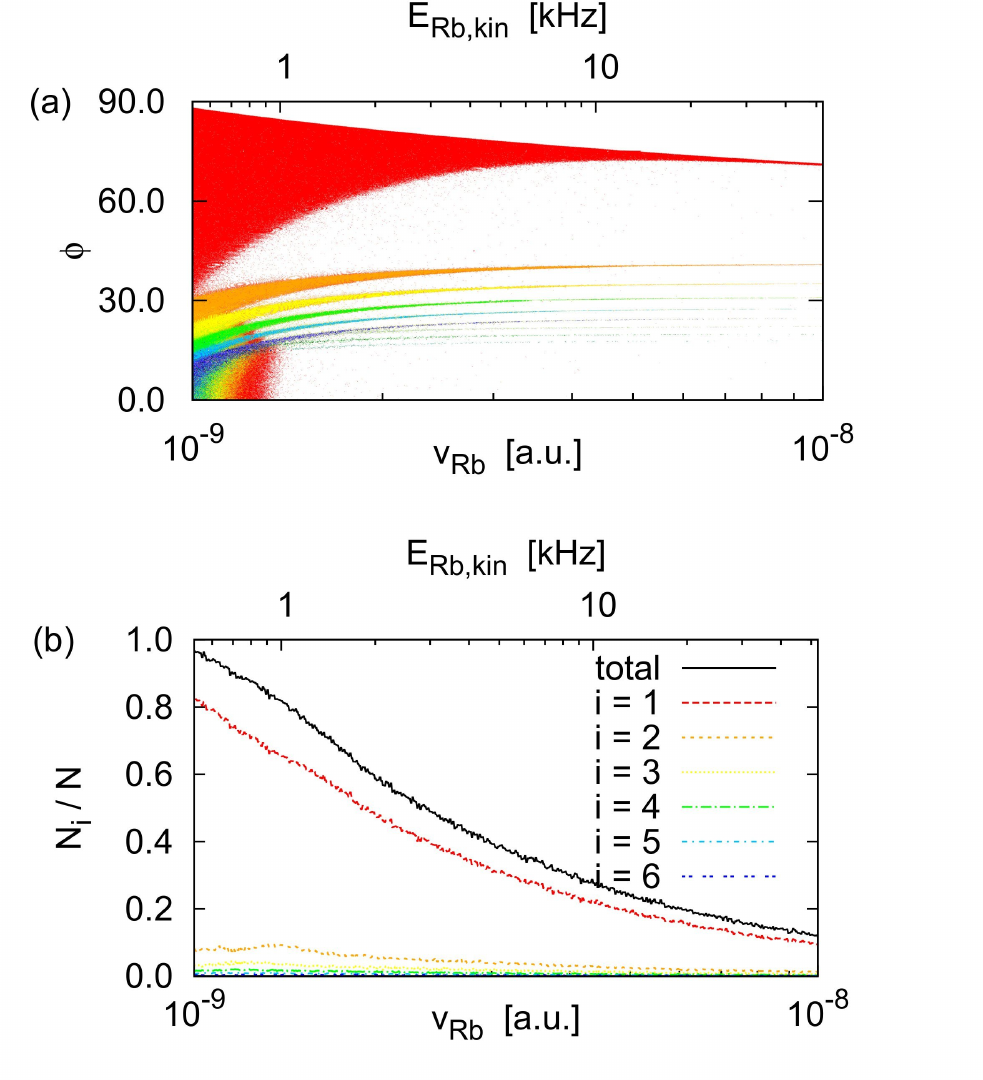}
\caption{\label{fig-capturing}
 (a) Result of the dynamics calculations of the \GSA\ for initial values
 $R_0=2500\,$a.u.,  $v_\mathrm{Rb}= 10^{-9}$ to
 $v_\mathrm{Rb}= 10^{-8}$ and angles $\phi$ from $0^\circ$ to
 $90^\circ$. Each single point represents a set of initial conditions
 ($v_\mathrm{Rb}$,$\phi$) and its color indicates in which minimum of
 the potential the \GSA\ will come to rest (colors refer to Fig.\
 \ref{fig-potential}). White means that there is no capturing, i.e.,
 the \GSA\ will again leave the Rydberg atom.
 (b) Fraction $N_i/N$ of the $N_i$ \GSA s that are captured in the $i$-th
 potential minimum out of a total of  $N$ atoms as a function of their
 velocity $v_\mathrm{Rb}$. The solid line shows the total fraction of
 captured \GSA s.}
\end{figure}

Fig.\ \ref{fig-capturing}(a) shows the domains of initial values $\phi$
and $v_\mathrm{Rb}$ where capture occurs. The colors indicate in which
of the local potential minima the \GSA\ is captured, and the color
assignment is the same as in Fig.\ \ref{fig-potential}. It can be seen
from Fig.\ \ref{fig-capturing}(a) that it strongly depends on the initial
conditions whether or not the \GSA\ is captured. For \AZ{slow} atoms
there is a broad range of angles $\phi$ in which a molecule is formed
while this range quickly shrinks with increasing velocity of the
\GSA. The areas with branches reaching to $v_\mathrm{Rb}\gtrsim
10^{-8}\,$a.u.\ correspond to situations where the \GSA\ is
directly captured when approaching the Rydberg atom. For small angles
$\phi \lesssim 20^\circ$ and velocities $v_\mathrm{Rb}\lesssim
1.5\times 10^{-9}\,$a.u.\ we also find situations where the \GSA\ is
captured on its way out from the Rydberg atom after a reflection at
some minimum internuclear distance. For $v_\mathrm{Rb}\lesssim 
10^{-9}\,$a.u.\ (not shown) almost all angles $\phi$ lead to
capture. However, initial velocities of $v_\mathrm{Rb} \gtrsim
0.5\bar{v}$ corresponding to energies above the natural line width of
the Rydberg molecule would lead to off-resonant absorption and are
thus forbidden.

To compute probabilities with which a \GSA\ of the velocity
$v_\mathrm{Rb}$ is captured in the $i$-th potential minimum, we regard
a uniform current of $N$ \GSA s and a diameter of at least the
extension of the Rydberg atom interacting with it. Considering that
the different angles then occur with a weighting factor of $4\pi
\sin^2\phi$, integrating over the cross-section of this current and
dividing by all $N$ \GSA s yields the probability of being captured
in the $i$-th minimum. This fraction $N_i/N$ is shown in Fig.\
\ref{fig-capturing}(b) (colors again correspond to those in Fig.\
\ref{fig-potential}). As can be seen, the predominant part will be
captured in the outermost minimum ($i=1$), in which also the
vibrational ground state of the Rydberg molecule is located (see
Refs. \cite{Greene2000,Bendkowsky2009}). Capturing in other minima
also happens, but by far more rarely.

\subsection{Extended model including p-wave scattering}
\label{sec:pwave}
\begin{figure}[t]
\includegraphics[width=\columnwidth]{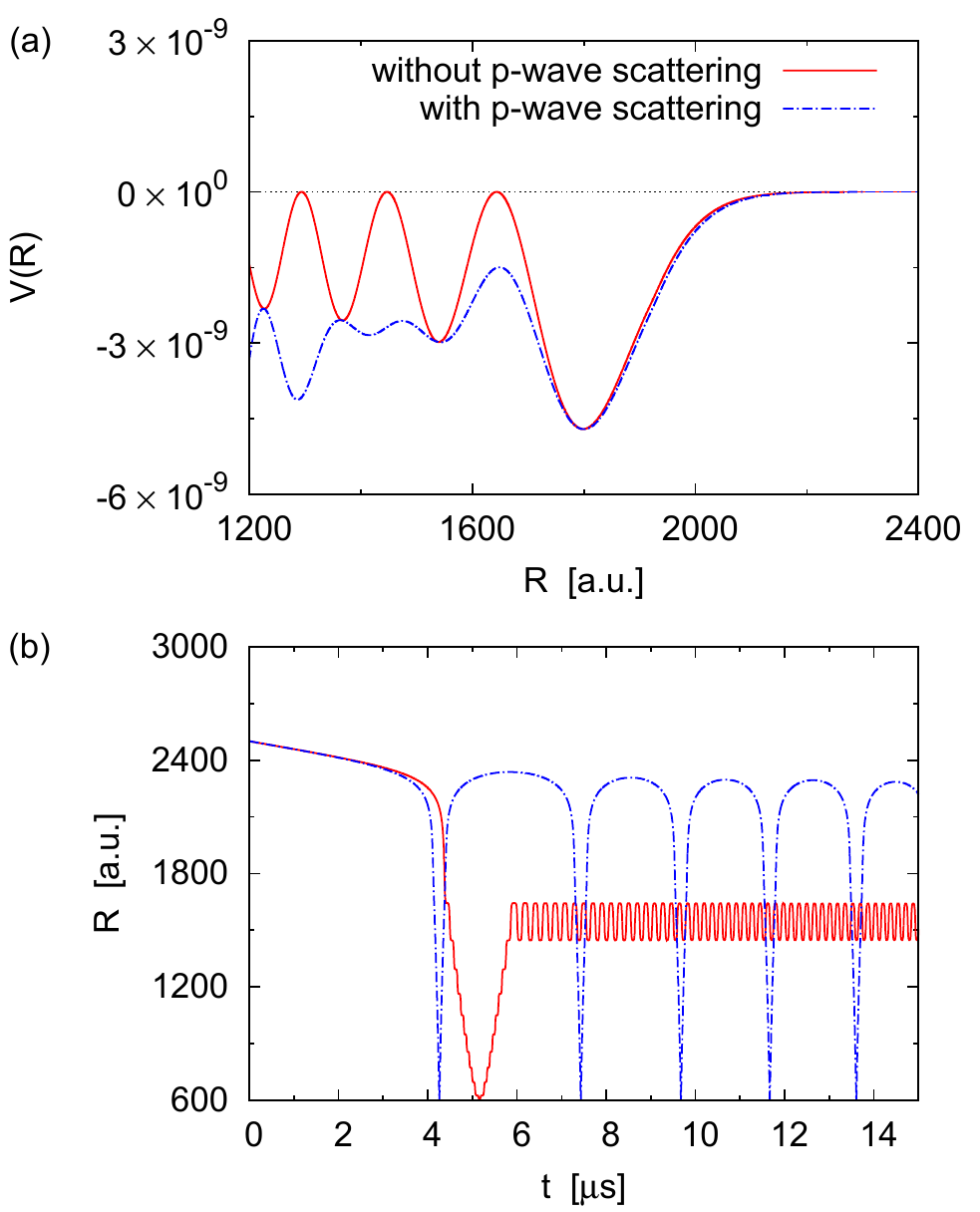}
\caption{\label{fig-p-wave}
  (a) Comparison of the molecular potential $V(\rr)$ with (solid
  line) and without (dashed line) p-wave scattering.
  (b) Comparison of two trajectories of the \GSA\ in the influence of
  the Rydberg atom for the same initial conditions with (dashed-dotted
  line) and without (solid line) p-wave
  scattering. The detailed dynamics change significantly, the fact
  \emph{that} the \GSA\ is captured, however, is not affected.}
\end{figure}

The potential \eqref{eq-potential} entering the equation of motion
\eqref{eq-DGL} is significantly modified near the Rydberg core when
p-wave scattering \cite{Bendkowsky2010, Khuskivadze2002} is also
taken into account by the additional term given in Eq.\ \eqref{eq-V-p}.
The main effect for the calculations performed in this paper is the
fact that the single potential wells are no more separated by some
internuclear separation $R$ with $V(R)=0$ [see Fig.\ \ref{fig-p-wave}(a)]. 
For a particle with $E \lesssim 0$ one is therefore no more able to
distinguish the different regions of capturing which, in the
quantum-classical model, is expressed by the fact that the reversal
points of the \GSA 's trajectory are shifted within the interaction
region: The outer reversal point will then in general be determined by
the exponential tail of the wave function ($R \sim 2400$) and the
inner one is given by the molecular potential together with the
centrifugal barrier for non-central collisions. However, also with 
p-wave scattering the potential remains conservative
so that the dissipative interaction described by the second term in
Eq.\ \eqref{eq-DGL} is not affected.

To illustrate the effects of p-wave scattering, Fig.\
\ref{fig-p-wave}(b) shows a comparison of two trajectories of a slow
\GSA\ one of which is obtained by taking account of p-wave scattering
(dashed-dotted line)  in the potential (\ref{eq-potential}) and the
other trajectory (solid line) by neglecting this effect. In the first
case the \GSA\ is no more captured in a particular well, however,
the point \emph{that} the \GSA\ is captured remains valid since in
both cases we have $V(R)\to 0$ for $R\to \infty$ which precludes a
\GSA\ with $E<0$ to escape from the Rydberg atom.

\subsection{Beyond the quantum-classical approach}
\label{sec:beyond-quantum-classical}
In the quasi-classical model the motion of the heavy ground state
atom is described purely classical. Nevertheless, the question
arises, what is the quantum state of the ground state atom when it
has been decelerated to energy $E<0\,$? The atom can no longer be in a
pure continuum state, however, the density of states at $E \lesssim
0$ is small and the energy gap to the highest vibrational bound
state is still rather large. The atom must therefore be in a
non-stationary superposition of eigenstates with an admixture of one
or more vibrational bound states. As explained above, the Rydberg
state of the electron must absorb the transferred energy (within the
natural line width of the state), and thus is slightly detuned. This
somehow unstable configuration may now develop in one of the
following directions.

On the one hand, the detuning of the Rydberg state may just lead
to a decrease of its lifetime. Reduced lifetimes of Rydberg
molecules in vibrational ground and excited states have already been
measured experimentally \cite{Butscher2011} but the mechanism for
that reduction has not yet been clarified. The dissipative finite
mass correction term which couples the Rydberg state and the motion
of the ground state atom may thus provide a physical interpretation
for the reduced lifetimes of Rydberg molecules.

On the other hand, the admixture of lower vibrational states to
the quantum dynamics of the ground state atom may enable the ground
state atom to jump into the next lower stationary vibrational state
by photon emission with a much higher probability than is to be
expected for \emph{spontaneous} emission, which is negligibly small
due to the small energy differences between the states. The detailed
investigation of such a process, which would require the treatment
within quantum electrodynamics, however, goes far beyond the scope
of this paper.

\section{Conclusion and outlook}
Investigating the interaction between a Rydberg electron and a \GSA\
in Rydberg excited gases within a quantum-classical framework, we were
able to derive a dissipative finite-mass correction term to the
classical equations of motion describing the dynamics of a \GSA\
interacting with a Rydberg atom.
Considering this correction term of order $\order{\me/\mRb}$ we have
shown that a free \GSA\ can, for suitable initial conditions, be
captured by the Rydberg atom and thus form a Rydberg
molecule. According to our calculations, this process takes place for 
``slow'' \GSA s, and 
does not depend on whether or not p-wave scattering is considered
in the molecular potential. However, the classical paths with and
without p-wave scattering differ: While with pure s-wave scattering
capturing is most likely in the outermost potential minimum, there
is no capturing in a particular well when p-wave scattering is
included.

In the experiment of Butscher \etal\ \cite{Butscher2011} molecular
Rb$_2^+$ ions have been detected in the photoassociation spectra at
resonance of the Rydberg atom, i.e., without any detuning of the
lasers. The formation of Rydberg molecules with the
quantum-classical model introduced in this paper and a subsequent
ionization process may provide a first hint towards a physical
explanation of that observation, however, further investigations are
necessary to get a deeper understanding of the dynamics of
ultralong-range Rydberg molecules.

%

\end{document}